\pdfoutput=1

\documentclass[12pt,preprint]{aastex62}
\usepackage{amsmath}
\usepackage{graphicx}
\usepackage{mathrsfs}
\usepackage{color}
\usepackage{url}
\usepackage{CJKutf8}
\usepackage{ulem}
\usepackage{wasysym}

\received{2021}
\revised{\today}
\accepted{\today}
\shorttitle{2020 XL$_{5}$}
\shortauthors{Hui et al. 2021}

\begin{document}

\title{
The Second Earth Trojan 2020 XL$_{5}$
}

\correspondingauthor{Man-To Hui}
\email{mthui@must.edu.mo}

\author{\begin{CJK}{UTF8}{bsmi}Man-To Hui (許文韜)\end{CJK}}
\affiliation{State Key Laboratory of Lunar and Planetary Science, 
Macau University of Science and Technology, 
Avenida Wai Long, Taipa, Macau}
%\nocollaboration

\author{Paul A. Wiegert}
\affiliation{Department of Physics and Astronomy, 
The University of Western Ontario, London, Ontario N6A 3K7, Canada}
\affiliation{Institute for Earth and Space Exploration, 
The University of Western Ontario, London, Ontario N6A 3K7, Canada}

\author{David J. Tholen}
\affiliation{Institute for Astronomy, University of Hawai`i, 
2680 Woodlawn Drive, Honolulu, HI 96822, USA}

\author{Dora F{\"o}hring}
%\affiliation{ESA NEO Coordination Centre, Largo Galileo Galilei,
%1, 00044 Frascati (RM), Italy}
\affiliation{Institute for Astronomy, University of Hawai`i, 
2680 Woodlawn Drive, Honolulu, HI 96822, USA}

%\author{Denise Hung}
%\affiliation{Institute for Astronomy, University of Hawai`i, 
%2680 Woodlawn Drive, Honolulu, HI 96822, USA}

%\nocollaboration

\begin{abstract}

The Earth Trojans are co-orbitals librating around the Lagrange points $L_4$ or $L_5$ of the Sun-Earth system. Although many numerical studies suggest that they can maintain their dynamical status and be stable on timescales up to a few tens of thousands of years or even longer, they remain an elusive population. Thus far only one transient member (2010 TK$_7$) has been discovered serendipitously. Here, we present a dynamical study of asteroid 2020 XL$_5$. With our meticulous followup astrometric observations of the object, we confirmed that it is a new Earth Trojan. However, its eccentric orbit brings it close encounters with Venus on a frequent basis. Based on our N-body integration, we found that the asteroid was captured into the current Earth Trojan status in the 15th century, and then it has a likelihood of 99.5\% to leave the $L_4$ region within the next $\sim$10 kyr. Therefore, it is most likely that 2020 XL$_5$ is dynamically unstable over this timescale.

\end{abstract}


\section{Introduction}
\label{sec_intro}

Trojans are small bodies in 1:1 mean-motion resonance (MMR) with some planet, librating around the Lagrange points $L_4$ or $L_5$ of the Sun-planet system. Numerous researchers \citep[e.g.,][and many more]{2002MNRAS.334..241B, 2021MNRAS.507.1640C, 2012MNRAS.426.3051C, 2013CeMDA.117...91M, 1992AJ....104.1641M, 1967AJ.....72...10R, 2000MNRAS.319...63T, 2019A&A...622A..97Z} have performed theoretical and numerical studies on dynamical stability of Trojans and other types of co-orbitals of solar system planets.

In the past century, over ten thousand Trojans have been recognised, the vast majority of which belong to Jupiter. In contrast, only one Earth Trojan (asteroid 2010 TK$_7$) has been identified, the discovery of which was in a serendipitous way by the Wide-field Infrared Survey Explorer mission in space \citep{2011Natur.475..481C}. Ironically, all dedicated Earth Trojan surveys have found nothing \citep[e.g.,][]{2021AJ....161..282L, 2020MNRAS.492.6105M, 1998Icar..136..154W}, owing to the fact that the observing circumstances of these objects are never ideal for ground-based telescopes, because they are always at small solar elongations. Moreover, such a survey will have to cover a huge sky area where Earth Trojans are potentially residing because of their proximity to Earth and libration around the Lagrange points \citep{2000Icar..145...33W}. 

Dynamical studies \citep[e.g.,][]{2013CeMDA.117...91M,2019A&A...622A..97Z} indicate that low-inclination Earth Trojans have the potential to survive the age of the solar system, and therefore these long-term stable members could be primordial planetesimal remnants formed in situ near the Earth-Moon system in the protoplanetary disc. However, none of this subgroup of Earth Trojans has been discovered yet. In comparison, Mars has several Trojans which are dynamically stable over the age of the solar system \citep[e.g.,][]{2013MNRAS.432L..31D,2005Icar..175..397S}. The only known Earth Trojan 2010 TK$_7$ is dynamically stable in the 1:1 MMR with Earth for merely $\la$25 Myr, and therefore it was most likely captured from elsewhere as a near-Earth asteroid \citep{2012A&A...541A.127D}.

Recently, asteroid 2020 XL$_5$ was discovered by the Panoramic Survey Telescope and Rapid Response System (Pan-STARRS 1) telescope at Haleakala Observatory, Hawai`i, on 2020 December 12,\footnote{\url{https://www.minorplanetcenter.net/mpec/K20/K20XH1.html}} and was suspected to be a potential candidate of the Earth Trojan population \citep{2021RNAAS...5...29D}. Thanks to our followup astrometric observations, we report our conclusive identification of asteroid 2020 XL$_{5}$ as the second transient Earth Trojan, confirming the study by \citet{2021RNAAS...5...29D} based upon a much shorter observing arc of the object. We describe our followup observations in Section \ref{sec_obs} and present a dynamical analysis of the asteroid in Section \ref{sec_dyn}.

\section{Observation}
\label{sec_obs}

We obtained followup observations of 2020 XL$_5$ using the University of Hawai`i (UH) 2.2 m telescope atop Mauna Kea, Hawai`i, on UT 2021 January 8, and September 12 and 13. In the first observing night, the observations were acquired through the Tektronix CCD in the on-chip $2 \times 2$ binning mode, rendering us an angular resolution of 0\farcs44 pixel$^{-1}$ with a field-of-view (FOV) of $7\farcm5 \times 7\farcm5$. In the remaining observing nights, we employed the STAcam CCD, which has been in use since April 2021. To achieve critically sampling the typical seeing on Mauna Kea, the STAcam images were $5 \times 5$ binned, resulting in a pixel scale of 0\farcs41 and an image dimension of $2112 \times 2112$ pixels. Individual images from 2021 January 8 have exposures of 60 and 90 s, while those from September 2021 have a common exposure of 8 min. In order to maximise collecting photons from the asteroid, we did not employ any filter, and the telescope was tracked nonsidereally at the apparent motion rate of the target. During our observations, the weather remained totally clear.

We performed astrometric measurements for our followup observations. Since our observations were tracked nonsidereally, all of the field stars in the same FOV alongside the target were obviously trailed in the data, making conventional simple centroiding techniques inapplicable. To accommodate this, we treated each star trail as a trapezoid and Gaussian in the along-track and cross-track directions, respectively. In such a profile model, there are six free parameters in total to be fitted using the least squares method, including the centroid's pixel coordinates, length, width, and position angle of the trail, and the peak pixel value of the trail profile model. The parameters were solved iteratively with determination of the sky background with adjacent pill-shaped annuli centred on the best-fit pixel coordinates of the centroids. With the obtained pixel coordinates, our code then used the least squares method to find the astrometric plate constants with the Gaia DR2 catalogue \citep{2018A&A...616A...1G}, whereby we could then convert the pixel coordinates of 2020 XL$_5$, whose image profile was simply treated as a bidimensional Gaussian, to the R.A. and decl. coordinates in the J2000 system. The astrometric measurement uncertainties were estimated via error propagation by assuming the Poisson statistics for the observing data. Our results are tabulated in Table \ref{tab:astrom}. We estimated the seeing using the full-width at half maximum (FWHM) of the asteroid as well as the FWHM of star trails in the cross-track direction, finding that the values varied between $\sim$0\farcs6 and 1\farcs0 from night to night.

\section{Dynamics}
\label{sec_dyn}

\begin{deluxetable*}{c|c|c|c|c|c|c}
%\rotate
\tablecaption{Our Followup Astrometric Observations of 2020 XL$_5$
\label{tab:astrom}}
\tablewidth{0pt}
\tablehead{
Observation Time  & R.A. ($^{\rm h}$ $^{\rm m}$ $^{\rm s}$) & decl. (\degr~\arcmin~\arcsec) & 
\multicolumn{2}{c|}{$1\sigma$ Uncertainty (\arcsec)} &
\multicolumn{2}{c}{O-C Residuals\tablenotemark{\dag} (\arcsec)}\\ \cline{4-7}
(UTC) & $\alpha$ & $\delta$ & E-W ($\Delta \alpha \cos \delta$) & decl. ($\Delta \delta$) & E-W & decl.
}
\startdata
2021 Jan 08.661324 & 14 33 49.544 & $-$19 59 19.23 & 0.075 & 0.075 & $+0.028$ & $-0.013$ \\
2021 Jan 08.666390 & 14 33 51.824 & $-$19 59 21.33 & 0.071 & 0.070 & $-0.020$ & $+0.015$ \\
2021 Sep 12.610790 & 07 07 42.372 & $+$09 42 55.34 & 0.066 & 0.066 & $-0.061$ & $-0.036$ \\
2021 Sep 12.616674 & 07 07 43.246 & $+$09 42 52.64 & 0.095 & 0.095 & $-0.039$ & $+0.073$ \\
2021 Sep 12.622738 & 07 07 44.143 & $+$09 42 49.72 & 0.066 & 0.066 & $-0.067$ & $+0.049$ \\
2021 Sep 13.617119 & 07 10 13.089 & $+$09 34 46.08 & 0.045 & 0.045 & $+0.018$ & $-0.019$ \\
2021 Sep 13.623080 & 07 10 13.974 & $+$09 34 43.20 & 0.054 & 0.054 & $+0.046$ & $+0.002$ \\
%\hline
\enddata
\tablenotetext{\dag}{Observed minus calculated residuals in our best-fit orbital solution in Table \ref{tab:orb}.}
\tablecomments{The coordinates are referred to the Earth mean equator and equinox of J2000 system. Technically the uncertainty and residuals in the R.A. direction are essentially in the east-west (E-W) direction on the corresponding great circle in the celestial sphere.}
\end{deluxetable*}

\begin{deluxetable}{lc|c}
%\rotate
\tablecaption{Our Best-Fit Orbital Solution for New Earth Trojan 2020 XL$_5$ (Heliocentric Ecliptic J2000.0)
\label{tab:orb}}
\tablewidth{0pt}
\tablehead{
\multicolumn{2}{c|}{Quantity}  & 
Value
}
\startdata
Semimajor axis (au) & $a$
       & 1.00076222(54)  \\ 
Eccentricity & $e$
       & 0.3871541(17)  \\ 
Inclination (\degr) & $i$
       & 13.846827(10)  \\ 
Longitude of ascending node (\degr) & ${\it \Omega}$
                 & 153.598852(87) \\ 
Argument of perihelion (\degr) & $\omega$
                 & 87.984486(98) \\ 
Mean Anomaly (\degr) & $M$
                  & 262.41609(13) \\
\hline
%\multicolumn{2}{l|}{Osculation epoch (TDB)} 
%& JD 2459276.5 = 2021 Mar 03.0
%& JD 2459276.5 = 2021 Mar 03.0 \\
%\multicolumn{2}{l|}{Observed arc}
%& 2021 Mar 02-03 (21.1 hr)
%& 2021 Mar 02-03 (21.1 hr) \\
%\multicolumn{2}{l|}{\# of observations}
%& 9
%& 9 \\
\multicolumn{2}{l|}{Weighted RMS residuals (\arcsec)}
& 0.112 \\
\multicolumn{2}{l|}{\# of observations}
& 27 \\
\multicolumn{2}{l|}{Observed arc}
& 2020 Nov 26-2021 Sep 13
\enddata
\tablecomments{The orbital elements are referred to an osculation epoch of TT 2020 November 26.5 = JD 2459180.0. The reported uncertainties (notated in the parenthesis fashion for concision) are all $1\sigma$ formal errors propagated from the astrometric measurement uncertainties.}
\end{deluxetable}

\begin{figure*}
\epsscale{1.0}
\begin{center}
\plotone{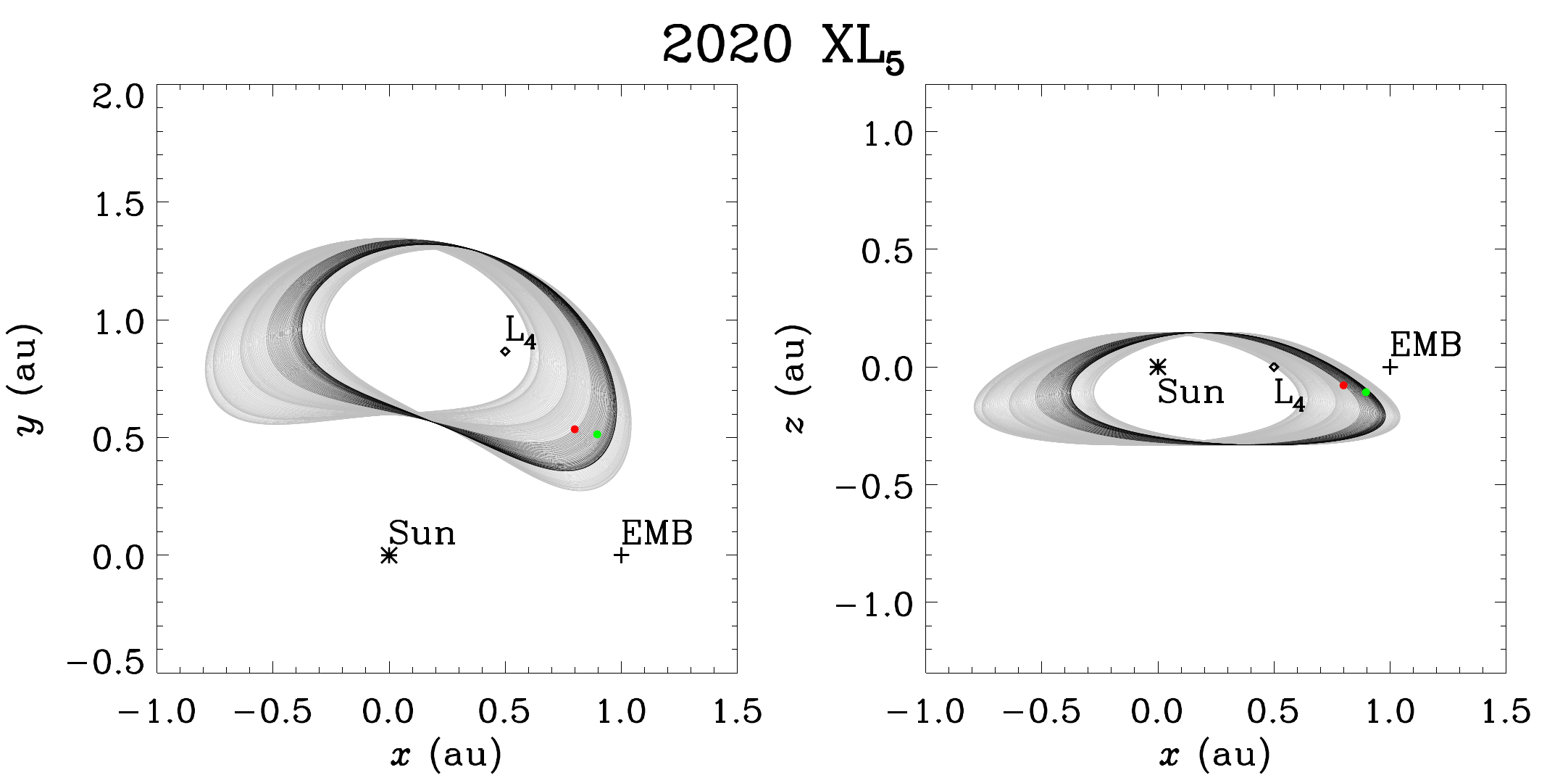}
\caption{
Trajectory of the new Earth Trojan 2020 XL$_5$ in the heliocentric frame co-rotating with the Earth-Moon barycentre (EMB, whose mean position is marked as the plus sign in the two panels) projected into the heliocentric orbital plane of the EMB (the $xy$ plane, left panel) and the $xz$ plane (right panel) from TDB 2021.0 to 2521.0. Also marked are the Sun (the asterisk) and the $L_4$ point. The red and green dots correspond to the starting and end points of the asteroid (moving clockwise in the $xy$ plane, and counterclockwise in the $xz$ plane). The section in black represents the trajectory of the asteroid in the first 50 yr of the timespan. Since the heliocentric orbit of the EMB is not circular but has a nonzero eccentricity, $e_{\rm EMB} \sim 0.01$, the epicycle of 2020 XL$_5$ shown here actually contains the contribution from the counterpart of the EMB, albeit much smaller.
\label{fig:crot}
} 
\end{center} 
\end{figure*}

\begin{figure*}
\epsscale{1.0}
\begin{center}
\plotone{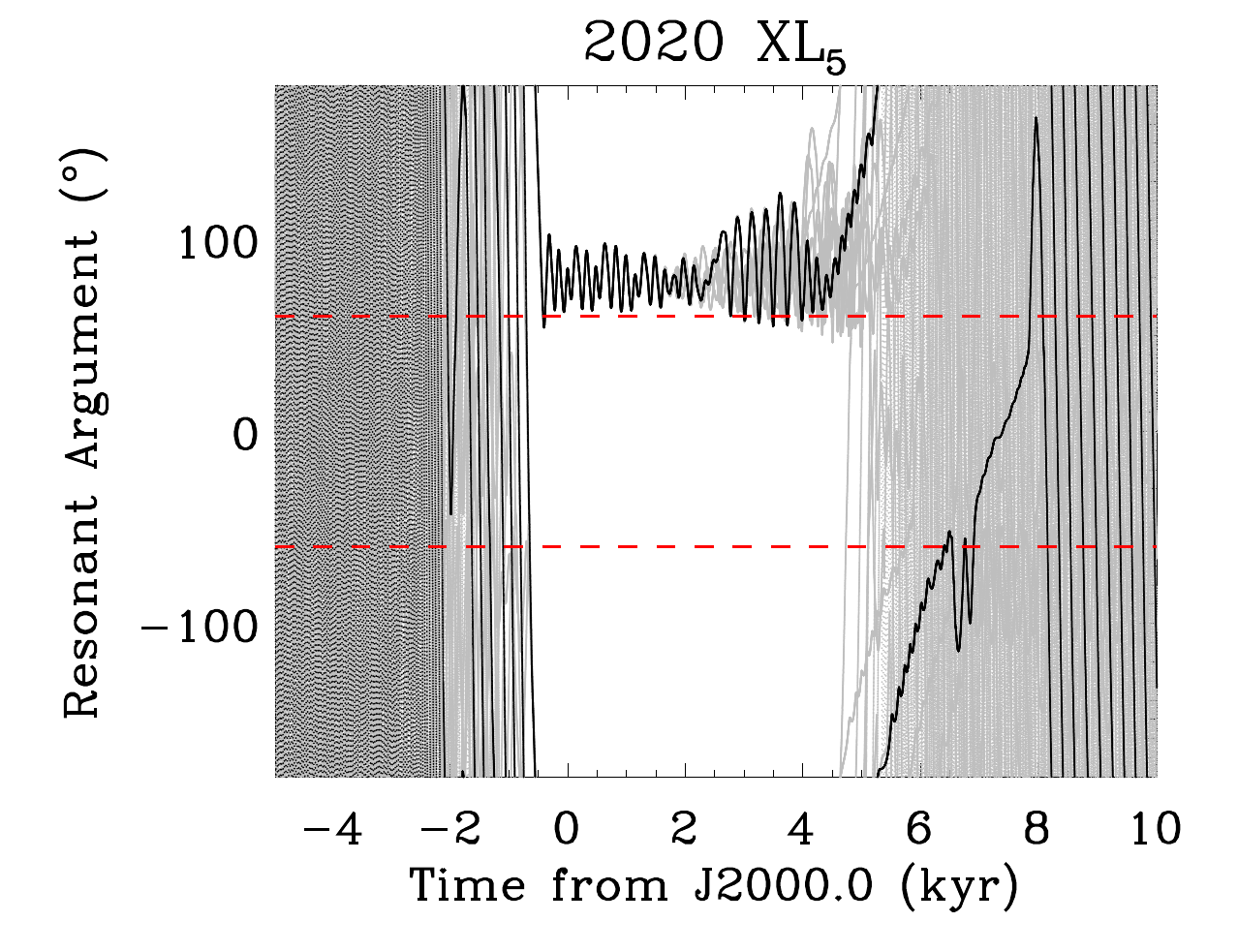}
\caption{
Temporal evolution of the resonant argument of asteroid 2020 XL$_5$ from BC $\sim$3000 to CE 12000 with every 0.2 yr marked. The evolution of the nominal orbit is in black, whereas the MC clones (only 10 out of 1000 shown for clarity) are in grey. The upper and lower horizontal straight lines in red correspond to the $L_4$ and $L_5$ points of the Sun-EMB system. Our numerical simulation suggests that the asteroid has been librating around the $L_4$ point since the 15th century, thereby being an $L_4$ Earth Trojan, and will maintain the current status for another few millennia with a libration period of $\sim$170 yr.
\label{fig:ra}
} 
\end{center} 
\end{figure*}

\begin{figure*}
\epsscale{1.0}
\begin{center}
\plotone{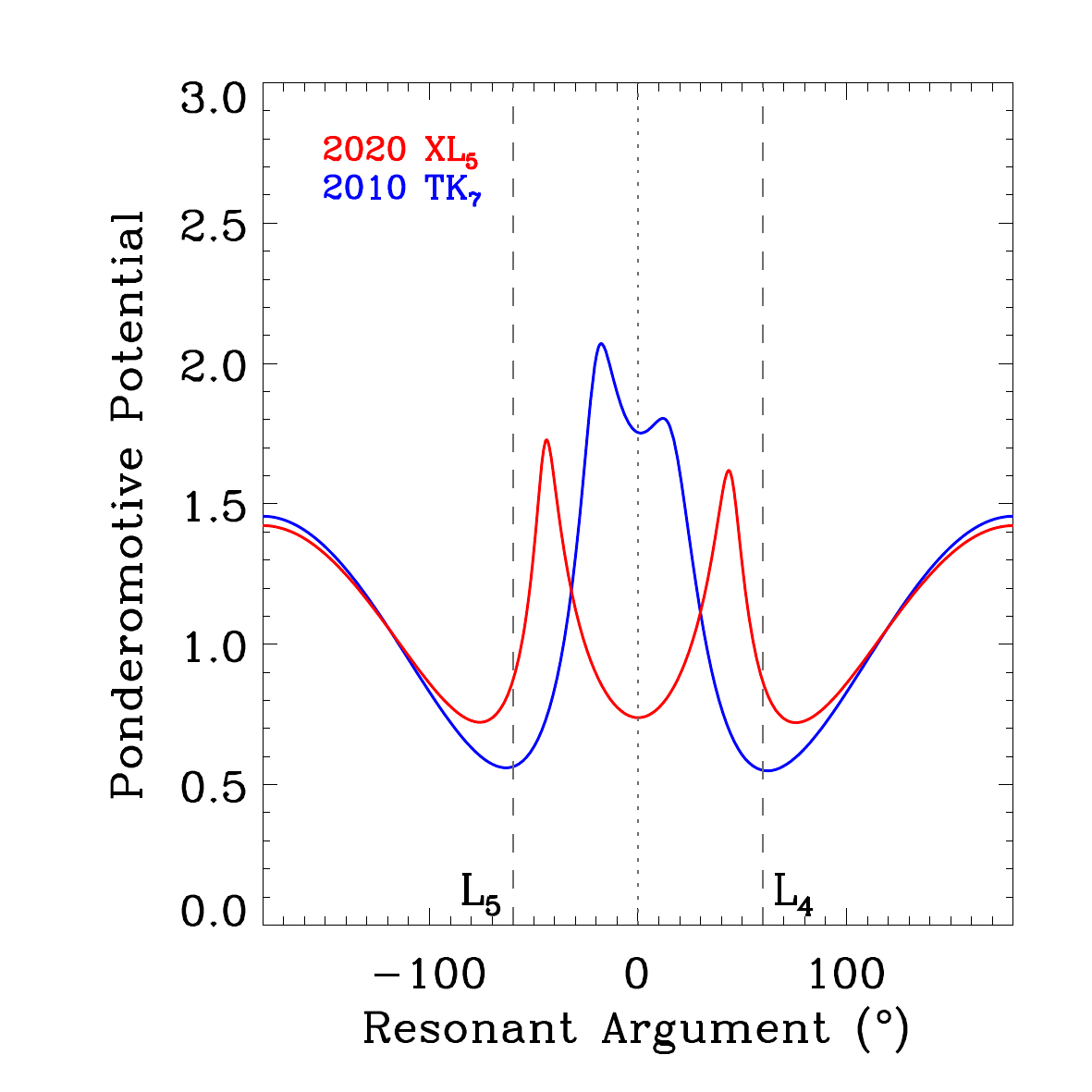}
\caption{
Comparison between the ponderomotive potentials of the two Earth Trojans 2020 XL$_5$ (red) and 2010 TK$_7$ (blue). Because of the nonzero inclination and eccentricity, minima of the effective potential are shifted away from $\pm60$\degr~in the resonant argument, respectively corresponding to the Lagrange points $L_4$ and $L_5$ (marked by two vertical dashed lines). The vertical dotted line marks the Sun-EMB direction.
\label{fig:S}
} 
\end{center} 
\end{figure*}

\begin{figure*}
\epsscale{1.0}
\begin{center}
\plotone{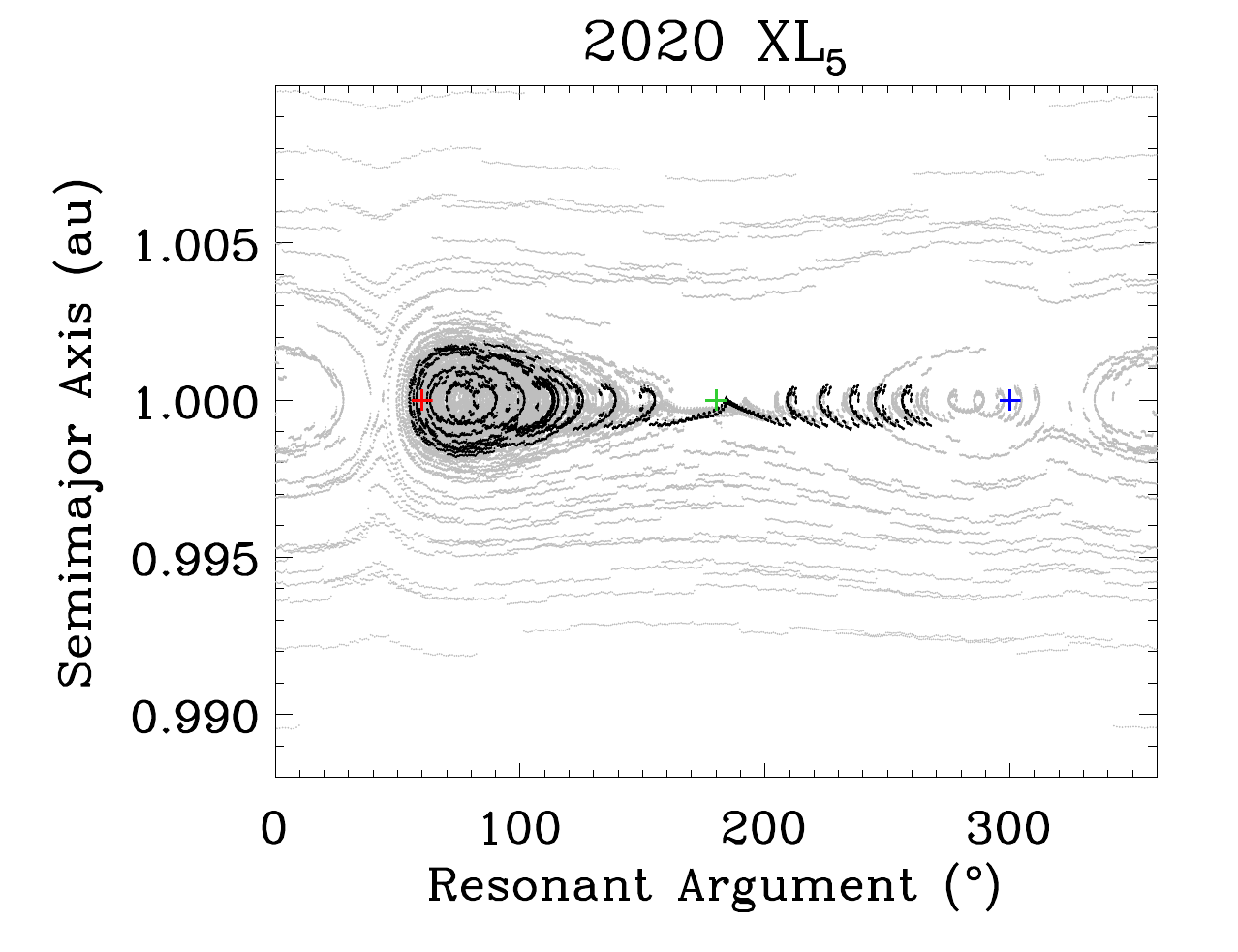}
\caption{
Phase portraits of 10 out of the 1,000 MC orbital clones (grey) and the nominal orbit (black) in a timespan from CE $\sim$5000 to 8000. The three crosses from left to right at the unity semimajor axis in au correspond to the three Lagrange points $L_4$, $L_3$, and $L_5$. Judging from the phase portraits, we can see that the while there are clones of 2020 XL$_5$ remaining as an $L_4$ Trojan, others may evolve into a transient $L_5$ Trojan, quasi-satellite, or may even leave the 1:1 MMR with the EMB. During this period, the nominal clone is gradually moving towards the $L_5$ point, as is evidenced in Figure \ref{fig:ra} as well.
\label{fig:phi_2}
} 
\end{center} 
\end{figure*}

\begin{figure*}
\epsscale{1.0}
\begin{center}
\plotone{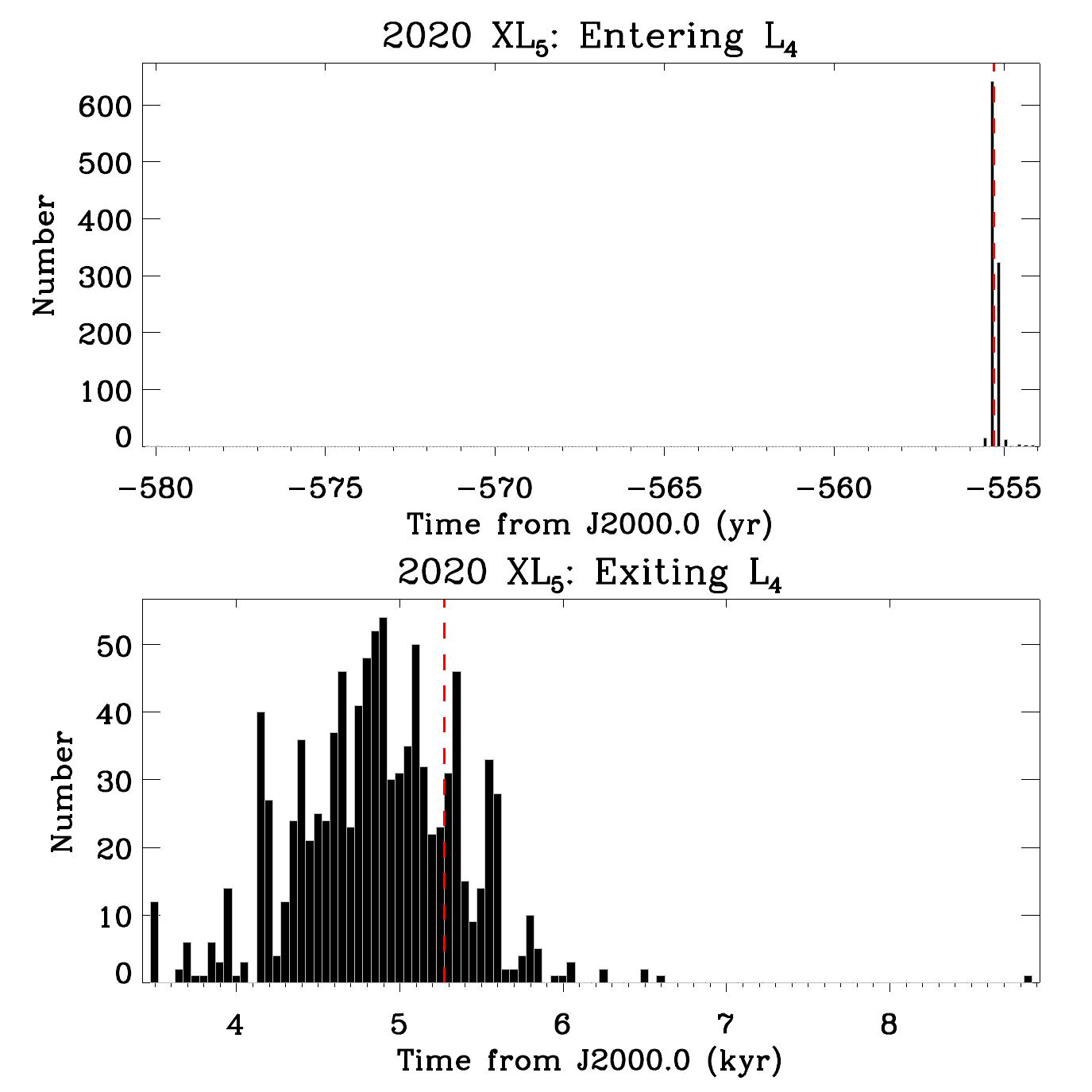}
\caption{
Statistics of the moments when the MC clones and the nominal orbit (red dashed line) entered (all of the clones) and leave (996 out of 1001 clones, or a fraction of 99.5\%) the Lagrange $L_4$ points of the Sun-Earth system.  Note that in the upper panel, there are two clones situated near the lower left corner, while the majority are distributed around CE $\sim$1445. See Section \ref{sec_dyn} for details.
\label{fig:stats_L4}
} 
\end{center} 
\end{figure*}

In order to investigate the dynamical status of 2020 XL$_5$, we updated its orbit with our astrometry together with astrometric measurements by other observers, including the Catalina Sky Surveys, Pan-STARRS 1, and a few others. These observations were obtained through querying the Minor Planet Center Database.\footnote{\url{https://www.minorplanetcenter.net/db_search}} Since there is no available information regarding the astrometric measurement uncertainties from these observers, we had to adopt the weighting scheme described in detail by \citet{2017Icar..296..139V} for them. Moreover, the observations were also debiased in accordance with \citet{2015Icar..245...94F}. We then refined the orbital elements  of 2020 XL$_5$ using the orbit determination package {\tt EXORB8} written by A. Vitagliano, in which the planetary and lunar ephemerides DE431 \citep{2014IPNPR.196C...1F} are utilised and perturbations from the eight major planets, Pluto, the Moon, and the 16 most massive asteroids as well as post-Newtonian corrections are all incorporated.\footnote{See a detailed description of the package at \url{http://www.solexorb.it/Solex120/Exorb81.pdf}.} We summarise our best-fit orbital elements for 2020 XL$_5$ as well as the associated $1\sigma$ formal errors, which were calculated from the obtained covariance matrix propagated from the astrometric measurement uncertainties, in Table \ref{tab:orb}. In the solution, all of the observations were found to have astrometric residuals within the assigned or measured error bars, indicating that our adoption of the weighting scheme is reasonable. We show the astrometric residuals of our followup observations in the best-fit solution in Table \ref{tab:astrom}. 

We then created 1,000 Monte Carlo (MC) orbital clones for 2020 XL$_5$ based on the obtained covariance matrix of the orbital elements according to the Cholesky decomposition method, which were subsequently integrated by {\tt SOLEX12}, an N-body integration standalone package companioned by {\tt EXORB8}. To keep consistency, the exact same force model was applied. We thereby obtained the geometric heliocentric state vectors of the nominal orbit, orbital clones, major planets, the Moon, Pluto, and the 16 most massive asteroids referenced to the J2000 ecliptic as functions of different epochs. Since here we only care about whether 2020 XL$_5$ is a new Earth Trojan or not, we first computed the heliocentric Cartesian coordinates of the target in a nonuniformly rotating frame in which the barycentre of the Earth-Moon system (EMB) always lies on the $x$-axis and the $z$-axis is defined by the normal of the heliocentric orbital plane of the EMB:
\begin{align}
\begin{pmatrix}
 x \\ 
 y \\ 
 z 
\end{pmatrix}
& =
\begin{pmatrix}
{\bf r} \cdot \hat{\bf r}_{\rm EMB} \\ 
{\bf r} - \left({\bf r} \cdot \hat{\bf n}_{\rm EMB} \right) \hat{\bf n}_{\rm EMB} - \left({\bf r} \cdot \hat{\bf r}_{\rm EMB} \right) \hat{\bf r}_{\rm EMB}\\ 
\left({\bf r} \cdot \hat{\bf n}_{\rm EMB} \right) \hat{\bf n}_{\rm EMB}
\end{pmatrix}
\label{eq_xrot}.
\end{align}
\noindent Here, ${\bf r} = \left(X,Y,Z \right)^{\rm T}$ is the heliocentric position vector of 2020 XL$_5$, and $\hat{\bf r}_{\rm EMB}$ and $\hat{\bf n}_{\rm EMB}$ are respectively the unit radial vector and normal of the heliocentric orbital plane of the EMB, all expressed in the heliocentric J2000 eclitpic reference frame. In the rotating reference frame, $L_4$ and $L_5$ points are at $\left(1/2, \sqrt{3}/2, 0\right)^{\rm T}$ and $\left(1/2, -\sqrt{3}/2, 0\right)^{\rm T}$ au, respectively. We visually tracked the motion of the nominal and MC clones both backward and forward for 500 yr with a time step of 1 d in the $xy$-plane of the rotating reference frame, finding that all of them are moving around the $L_4$ point, which is indicative of 2020 XL$_5$ being a potential Earth Trojan (Figure \ref{fig:crot}). 

However, such visualisation alone does not provide us with any conclusive answer, because other co-orbitals may behave similarly. So next, we converted the state vectors in the heliocentric J2000 ecliptic reference frame to heliocentric osculating orbital elements and computed the resonant argument in the 1:1 MMR configuration
\begin{align}
\nonumber
\varphi & \equiv \lambda - \lambda_{\rm EMB} \\
& = \left(\varpi + M \right) - \left(\varpi_{\rm EMB} + M_{\rm EMB} \right)
\label{eq_ra},
\end{align}
\noindent where $\lambda$, $\varpi$, and $M$ are respectively the mean longitude, longitude of perihelion, and mean anomaly of the asteroid, and the quantities with the subscript ``EMB'' refer to those of the EMB system. For Trojans around the $L_4$ point their resonant arguments oscillate around 60\degr, and those around the $L_5$ point they will librate around $\varphi = -60$\degr, whereas other types of co-orbitals have $\varphi$ oscillating around 0\degr~if they are quasi-satellites, or around 180\degr~for horseshoe co-orbitals. We show the evolution of the resonant argument of ten of the clones alongside the nominal orbit from 5 kyr in the past to 10 kyr in the future in Figure \ref{fig:ra}, where we can see that, since the 15th century or thereabouts, 2020 XL$_5$ has been in the current dynamical status. Its resonant argument indeed oscillates close to but not about $\varphi = 60$\degr~in a libration period of $\sim$170 yr. We calculated the libration period using the formula by \citet{2000ssd..book.....M}
\begin{equation}
T_{\rm L} = \frac{4\pi}{3}\sqrt{\frac{a_{\rm EMB}^{3}}{3 G \mathcal{M}_{\rm EMB}}}
\label{eq_TL}.
\end{equation}
\noindent Here, $a_{\rm EMB} \approx 1$ au and $\mathcal{M}_{\rm EMB} \approx 6 \times 10^{24}$ are respectively the semimajor axis of the heliocentric orbit and total mass of the EMB system, and $G = 6.67 \times 10^{-11}$ m$^3$ kg$^{-1}$ s$^{-2}$ is the gravitational constant. Inserting numbers, we found $T_{\rm L} \approx 220$ yr, which is slightly longer than the observed period.

In order to understand the reason why the current libration of asteroid 2020 XL$_5$ is not about $\varphi = 60$\degr, we calculated its ponderomotive (effective) potential \citep{1999PhRvL..83.2506N}
\begin{equation}
\Psi = \frac{1}{2\pi} \int\limits_{-\pi}^{\pi} \left(\frac{1}{\left|{\bf r} - {\bf r}_{\rm EMB}\right|} - {\bf r} \cdot {\bf r}_{\rm EMB} \right) {\rm d} \lambda %\Bigg|_{\Delta a = 0}
\label{eq_psi}.
\end{equation}
\noindent where the position vectors are both expressed in au. Using the osculating orbital elements in Table \ref{tab:orb}, we plot the effective potential of the asteroid as the red curve in Figure \ref{fig:S}, in comparison to that of the first Earth Trojan 2010 TK$_7$ in blue, whereupon we can observe that, while the minima of the latter are located basically around the Lagrange points $L_4$ and $L_5$, those of 2020 XL$_5$ are clearly shifted to larger resonant arguments in magnitude, due to the nontrivial orbital inclination and eccentricity. Therefore, we are now confident that, 2020 XL$_5$ is currently in a tadpole orbit librating near the $L_4$ point, thus being the second known Earth Trojan after 2010 TK$_7$. 

We proceed to investigate the orbital stability of 2020 XL$_5$. Unfortunately, chaos in its orbit prevents us from finding a conclusive answer pertinent to the exact history of 2020 XL$_5$ before its current dynamical status as an $L_4$ Earth Trojan. We calculated the Lyapunov timescale of the asteroid to be only a few hundred years by means of the tangent map method by \citet{1999CeMDA..74...59M}. Therefore, statistics from backward integration of the clones for timespans much longer than the Lyapunov timescale are physically meaningless, because this will cause a manifest increase in entropy of the system with time reversal, which clearly violates the second law of thermodynamics. Indeed, orbital evolution of the MC clones and the nominal orbit in our backward integration before CE $\sim$1000 is drastically different (for instance, in terms of the resonant argument, see Figure \ref{fig:ra}).

\begin{deluxetable*}{c|c|c|c}
%\rotate
\tablecaption{Close Encounters of 2020 XL$_5$ from TDB 1900 January 01 to 2200 January 01
\label{tab:ca}}
\tablewidth{0pt}
\tablehead{
Time (TDB)\tablenotemark{\dag} & Body & Close Approach Distance (au) & Relative Speed (km s$^{-1}$)
}
\startdata
1979 Feb $01.8877 \pm 0.0014$ & Venus & $0.094141 \pm 0.000013$ & $11.23909 \pm 0.00048$ \\
1987 Jan $28.6599 \pm 0.0048$ & Venus & $0.054339 \pm 0.000065$ & $12.7084 \pm 0.0028$ \\
1995 Jan $27.4661 \pm 0.0066$ & Venus & $0.050302 \pm 0.000090$ & $12.8850 \pm 0.0040$ \\
2003 Jan $29.0713 \pm 0.0081$ & Venus & $0.082103 \pm 0.000092$ & $11.6470 \pm 0.0033$ \\ 
2054 Feb $12.7629 \pm 0.0065$ & Venus & $0.06122 \pm 0.00013$ & $16.7320 \pm 0.0060$ \\ 
2062 Feb $12.5494 \pm 0.0032$ & Venus & $0.049010 \pm 0.000058$ & $16.1344 \pm 0.0028$ \\ 
2070 Feb $10.5425 \pm 0.0020$ & Venus & $0.070863 \pm 0.000041$ & $17.1707 \pm 0.0020$ \\ 
2179 Apr $04.878 \pm 0.014$ & Venus & $0.098895 \pm 0.000022$ & $13.891 \pm 0.017$ \\ 
2190 Jan $30.540 \pm 0.023$ & Venus & $0.04856 \pm 0.00040$ & $16.001 \pm 0.021$ \\ 
2198 Feb $01.512 \pm 0.018$ & Venus & $0.03767 \pm 0.00019$ & $13.676 \pm 0.013$ \\ 
\enddata
\tablenotetext{\dag}{The corresponding uncertainties are in days.}
\tablecomments{Our search for close encounters between asteroid 2020 XL$_5$ and the massive bodies in our force model was confined to a maximum close approach distance of 0.1 au. Under such configuration, Venus is the only planet that showed up in our search. The reported uncertainties are all standard deviations computed from the 1,000 MC clones plus the nominal orbit.}
\end{deluxetable*}

It is most likely that 2020 XL$_5$ will maintain its current Earth Trojan state for the next four millennia, whereafter its evolution becomes unclear again due to chaos in its orbit. Following \citet{2011Natur.475..481C} and \citet{2012A&A...541A.127D}, we visually examined the phase portraits of the nominal orbit and the MC orbital clones of 2020 XL$_5$ in the semimajor axis versus resonant argument space. Here, for clarity, only ten of the MC clones alongside the nominal orbit are shown in Figure \ref{fig:phi_2}, where we can observe that while some of the clones remain in the current dynamical status as an $L_4$ Earth Trojan, others may become an $L_5$ Earth Trojan, quasi-satellite, or even leave the 1:1 MMR with the EMB.

In order to better characterise the timescale on which 2020 XL$_5$ remains as an Earth Trojan at the $L_4$ point, we tracked the nominal orbit and its 1,000 MC orbital clones at each output time step (0.2 yr) in our numerical integration. In the forward simulation, whenever the resonant argument of a clone exceeds 180\degr~for the very first time, we marked the corresponding epoch as the moment at which it has left the $L_4$ point. In the backward simulation, the latest epoch at which the clone has $\varphi \ge 180\degr$ was treated as the time when it started to be trapped in the $L_4$ region. The results are plotted in Figure \ref{fig:stats_L4}. We found that 2020 XL$_5$ has been an $L_4$ Earth Trojan since year CE $1444.7 \pm 1.1$, and $\sim$99.5\% of the clones (996 out of 1001 clones, including the nominal orbit) will leave the $L_4$ libration region within the next 10 kyr. We found the mean epoch when the clones exit the $L_4$ region to be $4.86 \pm 0.51$ kyr from J2000, where the uncertainty is the standard deviation. Therefore, it is highly likely that the 2020 XL$_5$ is dynamically unstable over the investigated period ($\sim$10 kyr), thus being the second transient Earth Trojan after 2010 TK$_7$. The latter became an $L_4$ Earth Trojan $\sim$2 kyr ago and will maintain its current dynamical status for $\sim$15 kyr before becoming a horseshoe co-orbital or jumping into the neighbourhood of the $L_5$ point \citep{2012A&A...541A.127D}.

The short lifetime of 2020 XL$_5$ being on an Earth Trojan orbit is not unexpected, because its nontrivial eccentricity plus the relatively small orbital inclination makes it susceptible to the gravitational pulls of other terrestrial planets, in particular Venus, with which close encounters were found to occur on a common basis. Our finding agrees with the analysis by \citet{2021RNAAS...5...29D} based upon a much shorter observed arc of the object. With {\tt SOLEX12} we made use of the 1,000 MC orbital clones in addition to the nominal orbit and searched for close encounters between the asteroid and various massive bodies in our force model described earlier in this section having mutual close approach distances $d_{\min} \le 0.1$ au from TDB 1900 January 1.0 to 2200 January 1.0. The results are summarised in Table \ref{tab:ca}, where we can see that in the examined three centuries, Venus is the only massive body that had and will continue having ten close encounters with the Earth Trojan at mutual distances $\le$0.1 au in our search. 

In the following, we use order-of-magnitude calculation to estimate the timescale on which 2020 XL$_5$ will leave the 1:1 MMR with Earth solely due to the perturbation from Venus. We only consider close encounters between the two bodies at $d_{\min} \le 0.1$ au. During one such close approach to Venus, the change in the orbital energy of 2020 XL$_5$ is due to the influence of the Venusian gravitational potential
\begin{equation}
\left| \Delta E \right| = G\frac{\mathcal{M}{_{\venus}} \mathcal{M}}{d_{\min}}
\label{eq_dE_v},
\end{equation}
\noindent where $\mathcal{M}{_{\venus}}$ and $\mathcal{M}$ are respectively the masses of Venus and 2020 XL$_5$, and $d_{\min} \le 0.1$ au is the close approach distance between the two bodies. In the two-body problem, the heliocentric orbital energy of 2020 XL$_5$ is given by
\begin{equation}
E = -G \frac{\mathcal{M}_{\odot} \mathcal{M}}{2a}
\label{eq_E_orb}.
\end{equation}
\noindent Here, $\mathcal{M}_{\odot}$ is the mass of the Sun. Differentiating both sides, and equating the change in the orbital energy to the one by Equation (\ref{eq_dE_v}), we find the change in the semimajor axis in the heliocentric orbit of 2020 XL$_5$ is
\begin{equation}
\Delta a = \frac{2a^2}{d_{\min}} \left(\frac{\mathcal{M}_{\venus}}{\mathcal{M}_{\odot}} \right)
\label{eq_da},
\end{equation}
\noindent whereby we obtain that after a close encounter with Venus at $d_{\min} \le 0.1$ au, the Earth Trojan experiences a change of $\ga$$5 \times 10^{-5}$ au in semimajor axis in his heliocentric orbit. We approximate that 2020 XL$_5$ will stop being trapped in the 1:1 MMR with Earth once its semimajor axis differs from the one of Earth by over the Hill radius of the latter, which is $R_{\rm H} \approx 0.01$ au. Despite we did not check close encounters between the asteroid and Venus outside the timespan from 1900 to 2200, we do not expect that the frequency of such encounters while 2020 XL$_5$ is still in the 1:1 MMR with Earth is significantly different. Accordingly, we estimate that a total number of $\la$200 such close encounters with Venus will accumulatively destabilise the orbit of the Earth Trojan and gradually nudge it outside the 1:1 MMR with Earth. Given the expected occurrence frequency, the whole process will take merely $\la$6 kyr, which agrees rather well with our N-body numerical simulation, demonstrating that Venus is the primary perturbation source that influence the co-orbital status of 2020 XL$_5$ with Earth.

Note that in our analysis, we have completely omitted the Yarkovsky effect of the Earth Trojan due to anisotropic solar heating. However, we argue that our conclusions are not likely altered considerably even if there is such an effect. Following \citet{2013Icar..224....1F} and \citet{2017AJ....153...80H}, we compute the expected drift rate in the semimajor axis of the heliocentric orbit of the asteroid by comparing to asteroid (101955) Bennu, in which way we will need a size estimate for 2020 XL$_5$. Using our photometry and that from the MPC, assuming a typical asteroidal phase slope of $G_{\alpha} = 0.15$ in the model by \citet{1989aste.conf..524B}, we find the absolute magnitude of Earth Trojan to be $H = 20.4 \pm 0.5$ in the {\it G} band of the Gaia DR2 catalogue. We further simply assume a typical geometric albedo of 0.1 for the asteroid, thus obtaining its nucleus radius to be $\left(1.6 \pm 0.4 \right) \times 10^2$ m. Accordingly, we find the change rate in the semimajor axis of 2020 XL$_5$ is then $\sim$$3 \times 10^{-3}$ au Myr$^{-1}$, which is by no means comparable to the corresponding change rate in the semimajor axis of the heliocentric orbit due to the perturbation by Venus. Therefore, we conclude that our omission of the Yarkovsky effect will not introduce any noticeable deviation from the reality.

%\clearpage
\section{Summary}
\label{sec_sum}

The key conclusions of our study are listed as follows:

\begin{enumerate}

\item With our followup astrometric observations, we confirmed that 2020 XL$_5$ is a new Earth Trojan after 2010 TK$_7$.

\item 2020 XL$_5$ is only a transient Earth Trojan, as it has been librating around the $L_4$ point only since the 15th century, and its orbit is unstable on a $\sim$10 kyr timescale primarily due to frequent close approaches to Venus at mutual distances of $\la$0.1 au. 

\item The minima of its current effective potential are clearly shifted away from 60\degr~to a larger angle in the resonant argument because of its nontrivial orbital inclination and eccentricity.

\end{enumerate}

\acknowledgements
{
We thank the anonymous referee for their prompt review and insightful comments and suggestions on our paper, and Aldo Vitagliano for making his excellent packages {\tt EXORB8} and {\tt SOLEX12} available for us to exploit. Observations from the University of Hawai`i 2.2 m telescope on the summit of Mauna Kea, which is a significant and sacred cultural site to the aboriginal people and in the Hawaiian culture, were made possible thanks to hard maintenance work by the day crew. It is also our honour to have the opportunity to obtain observations from the Hawaiian sanctuary.
}

\vspace{5mm}
%\facilities{}

\software{{\tt EXORB8}, IDL, {\tt SOLEX12}}

%\appendix

%\section{Tabulated Best-Fit Parameters}

%\clearpage

\end{document}